\documentclass{aa}

\usepackage{graphicx}
\usepackage{txfonts}
\usepackage{natbib}
\bibpunct{(}{)}{;}{a}{}{,} 

\begin{document} 

   \title{Planetary Radio Interferometry and Doppler Experiment (PRIDE) technique: A test case of the Mars Express Phobos fly-by.}

      \author{
          D.A.~Duev
          \inst{1,2,3}
          \and
          S.V.~Pogrebenko
          \inst{2}
          \and
          G.~Cim\`{o}
          \inst{2,4}
          \and
          G.~Molera Calv\'{e}s
          \inst{5,2}
          \and
          T.M.~Bocanegra~Baham\'{o}n
          \inst{6,2,7}
          \and
          L.I.~Gurvits
          \inst{2,6}
          \and
          M.M.~Kettenis
          \inst{2}
          \and
          J.~Kania
          \inst{8,2}
          \and
          V.~Tudose
          \inst{9}
          \and
          P.~Rosenblatt
          \inst{10}
          \and
          J.-C.~Marty
          \inst{11}
          \and
          V.~Lainey
          \inst{12}
          \and
          P.~de~Vicente
          \inst{13}
          \and
          J.~Quick
          \inst{14}
          \and
          M.~Nickola
          \inst{14}
          \and
          A.~Neidhardt
          \inst{15}
          \and
          G.~Kronschnabl 
          \inst{15}
          \and
          C. Ploetz
          \inst{15}
          \and
          R.~Haas
          \inst{16}
          \and
          M.~Lindqvist
          \inst{16}
          \and
          A.~Orlati
          \inst{17}
          \and
          A.V.~Ipatov 
          \inst{18}
          \and
          M.A.~Kharinov 
          \inst{18}
          \and
          A.G.~Mikhailov
          \inst{18}
          \and
          J.E.J.~Lovell
          \inst{19}
          \and
          J.N.~McCallum
          \inst{19}
          \and
          J. Stevens
          \inst{20}
          \and
          S.A.~Gulyaev 
          \inst{21}
          \and
          T.~Natush
          \inst{21}
          \and
          S.~Weston 
          \inst{21}
          \and
          W.H.~Wang
          \inst{7}
          \and
          B.~Xia
          \inst{7}
          \and
          W.J.~Yang
          \inst{22}
          \and
          L.-F.~Hao
          \inst{23}
          J.~Kallunki
          \inst{24}
          \and
          O.~Witasse
          \inst{25}
          }

   \institute{California Institute of Technology, 1200 E California Blvd, Pasadena, CA 91125, USA\\
              \email{duev@caltech.edu}
         \and
             Joint Institute for VLBI ERIC, P.O. Box 2, 7990 AA Dwingeloo,
             The Netherlands
         \and
             Sternberg Astronomical Institute, Lomonosov Moscow State University,
             Universitetsky av. 13, 119991 Moscow, Russia
       \and
             ASTRON, the Netherlands Institute for Radio Astronomy, Postbus 2, 7990 AA, Dwingeloo, The Netherlands
       \and
             Aalto University, School of Electrical Engineering, Department of Radio Science and Engineering, 02150 Espoo, Finland
        \and
            Department of Astrodynamics and Space Missions,
            Delft University of Technology, 2629 HS Delft, The Netherlands
        \and
            Shanghai Astronomical Observatory,
            80 Nandan Road, Shanghai 200030, China
       \and
            Carnegie Mellon University,
            5000 Forbes Ave, Pittsburgh, PA 15213, USA
       \and
            Institute for Space Sciences,
            Atomistilor 409, P.O. Box MG-23,
            Bucharest-Magurele RO-077125, Romania
       \and
            Royal Observatory of Belgium,
            Ringlaan 3, 1180 Brussels, Belgium
       \and
            CNES/GRGS, 
            OMP 14 avenue Edouard Belin 31400 Toulouse, France
       \and
            IMCCE, Observatoire de Paris, PSL Research University, CNRS-UMR8028 du CNRS, UPMC, Lille-1, 77 Av. Denfert-Rochereau, 75014, Paris, France
       \and
            Observatorio de Yebes (IGN),
            Apartado 148, 19180, Guadalajara, Spain
       \and
            Hartebeesthoek Radio Astronomy Observatory, 1740 Krugersdorp, South Africa
       \and
            Federal Agency for Cartography and Geodesy, Geodetic Observatory of Wettzell, 60598 Frankfurt Am Main, Germany
       \and
            Department of Earth and Space Sciences, Chalmers University of Technology, Onsala Space Observatory, 439 92 Onsala, Sweden
       \and
            National Institute for Astrophysics, Radio Astronomy Institute, Radio Observatory Medicina, 75500 Medicina, Italy
       \and 
            Institute of Applied Astronomy, Russian Academy of Sciences, 
            Kutuzova Embankment 10, 191187 Saint-Petersburg, Russia
       \and
            School of Physical Sciences, University of Tasmania, Private Bag 37, Hobart, 7001, Australia
       \and
            CSIRO Astronomy and Space Science, Australia Telescope National Facility, Narrabri NSW 2390, Australia
       \and
            Institute for Radio Astronomy and Space Research, Auckland University of Technology, 
            Private Bag 92006, Auckland 1142, New Zealand
       \and
            Xinjiang Astronomical Observatory, Chinese Academy of Sciences, 830011 Urumqi, PR China
       \and
            Yunnan Astronomical Observatory, Chinese Academy of Sciences, 650011 Kunming, PR China
       \and
            Mets\"ahovi Radio Observatory, Aalto University, 02540 Kylm\"al\"a, Finland
       \and
            European Space Agency, ESA/ESTEC Scientific Support Office, 2200AG Noordwijk, The Netherlands
            }

   \date{Received 7 May 2016 / accepted 31 May 2016}

 
  \abstract
   {The closest ever fly-by of the Martian moon Phobos, performed by the European Space Agency's Mars Express spacecraft, gives a unique opportunity to sharpen and test the Planetary Radio Interferometry and Doppler Experiments (PRIDE) technique in the interest of studying planet - satellite systems.}
   {The aim of this work is to demonstrate a technique of providing high precision positional and Doppler measurements of planetary spacecraft using the Mars Express spacecraft. The technique will be used in the framework of Planetary Radio Interferometry and Doppler Experiments in various planetary missions, in particular in fly-by mode.} 
   {We advanced a novel approach to spacecraft data processing using the techniques of Doppler and phase-referenced very long baseline interferometry spacecraft tracking.}
   {We achieved, on average, mHz precision ($30~\mathrm{\mu}$m/s at a 10 seconds integration time) for radial three-way Doppler estimates and sub-nanoradian precision for lateral position measurements, which in a linear measure (at a distance of 1.4~AU) corresponds to $\sim50$~m.}
   {}


   \maketitle
%

\section{Introduction}

The Planetary Radio Interferometry and Doppler Experiments (PRIDE) project \citep{2012A&A...541A..43D} initiated by the Joint Institute for VLBI ERIC (JIVE, Dwingeloo, The Netherlands) utilises Doppler and phase-referencing very long baseline interferometric (VLBI\footnote{Very long baseline interferometry (VLBI) is a technique in which the signals from a network of radio telescopes, spread around the world, are combined to simulate one single telescope with a resolution far surpassing that of the individual telescopes.}) observations to provide high precision spacecraft state vector estimation. In this paper, we further advance a novel approach to spacecraft data processing developed within PRIDE. 

On December 29, 2013, the European Space Agency's Mars Express (MEX) spacecraft made the closest ever fly-by of Phobos, one of the two Martian moons, just some 45 km from its surface \citep{2014P&SS..102...18W}. PRIDE observations of MEX during this event, involving more than 30 radio telescopes spread around the globe, were carried out by our team on December 28-29, 2013 (PI Pascal Rosenblatt, Royal Observatory of Belgium; European VLBI Network (EVN)/Global VLBI experiment code GR035). These observations allow reconstructing MEX's trajectory in the vicinity of Phobos with a high accuracy, which will in turn help to put a better constraint on the geophysical parameters of Phobos, possibly shedding light on its origin. The PRIDE data processing technique has been specifically refined for the observations of MEX during this event to provide high precision positional and Doppler measurements. In particular, we describe here the improvements made to the correlator software at JIVE \citep{2015ExA...tmp....9K} that allow efficient handling of such data, and we demonstrate the positive impact of these enhancements on the spacecraft position estimates obtained at the post-processing stage. The use of the Doppler and VLBI data in dynamical modelling of MEX motion to estimate the geophysical parameters related to the interior composition of this Mars moon is discussed in Rosenblatt et al. (2016, in prep.).

The GR035 experiment has been conducted as a live end-to-end verification of the PRIDE technique which will be used in future planetary missions, in particular ESA's Jupiter Icy Satellites Explorer, JUICE \citep{2013P&SS...78....1G}. In this paper, we present the technique, including the data processing algorithms. In Section 2, we describe the set-up of the GR035 experiment. Section 3 describes Doppler and VLBI data processing pipeline and presents the results of the experiment. Section 4 provides the reader with conclusions and outlook.
The scientific evaluation of the experiment will be given in a separate paper (Rosenblatt et al., in prep.). The experiment GR035 was complementary to the nominal MEX Radio Science experiment MaRS \citep{2004ESASP1240..141P,Paetzold201644}. 


\section{Experiment GR035 set-up}

The error in the a priori position of a phase calibrator directly affects the estimates of the MEX orbit parameters proportionally to the separation angle between the calibrator and the target. Therefore, in order to reach the best positional accuracy of MEX achievable with the modern ground-based VLBI -- about a nanoradian (0.2 mas), which is translated to $\sim$100 metres at the orbit, phase calibrators are necessary within $\sim$$2^{\mathrm{o}}$ \citep{1995ASPC...82..327B} of the MEX position at the observational epoch with absolute errors in their position not exceeding 0.2 mas.

\begin{figure}
   \centering
   \includegraphics[width=250pt,clip]{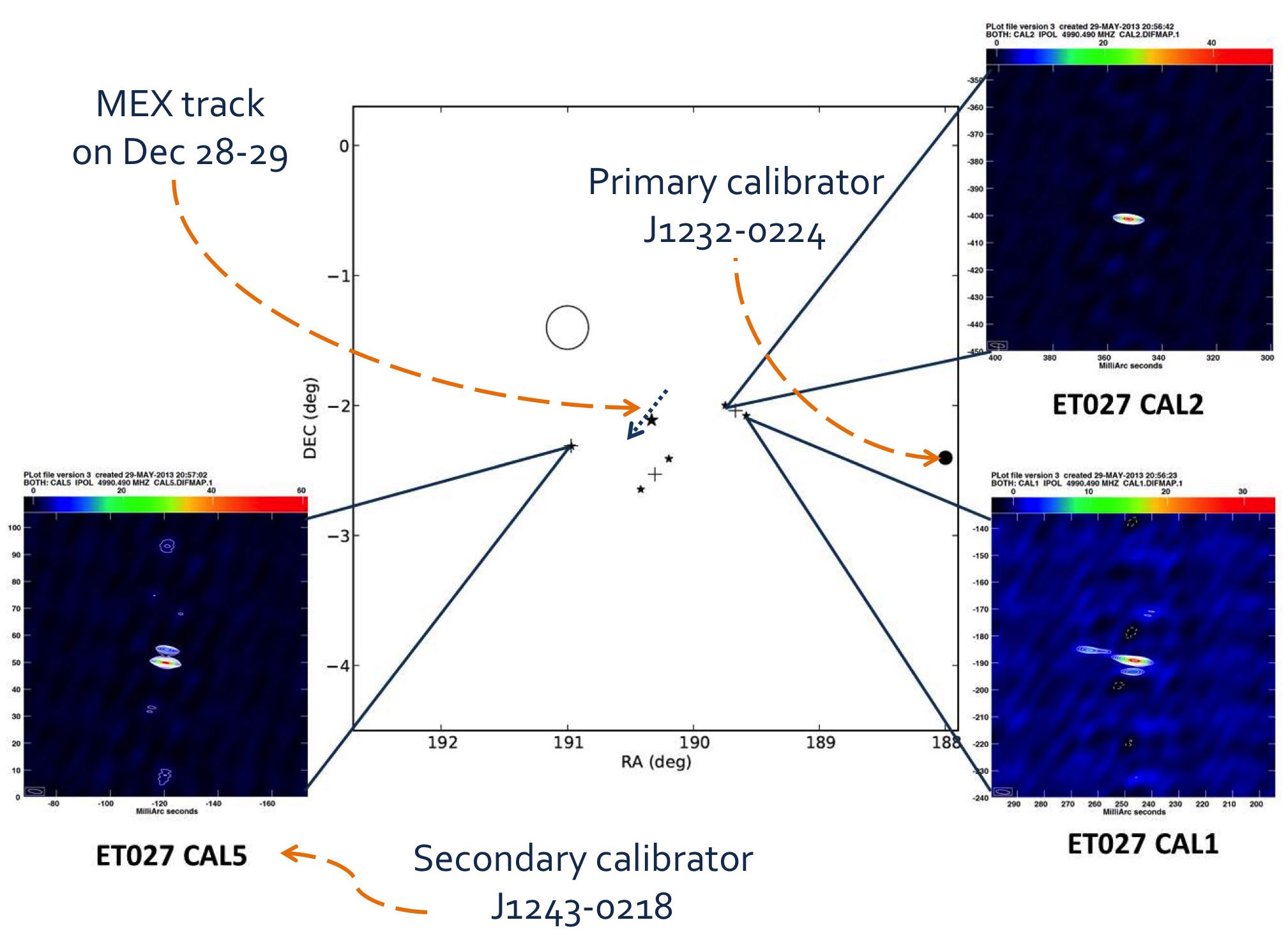}
   \caption{Mars Express Phobos fly-by experiment field on the sky. A large black star denotes the field centre; three crosses denote the pointings used in the experiment ET027 to observe possible secondary calibrators; five smaller black stars denote the detected sources; the circle represents the approximate size of the primary beam of a 30 m antenna. The colour insets show calibrated images of three of the sources in the field obtained in the ET027 experiment.}
   \label{gr035:setup}
    \end{figure}

In the case of the GR035 experiment, the timing of the observations and therefore the sky position of the events were, obviously, defined by MEX ballistics. The nearest bright source suitable as a primary phase calibrator, \object{J1232$-$0224}, with 0.16 mas error in RA, 0.27 mas error in Dec, and 0.717 Jy total X-band flux density \citep{RFC}, happened to be at a fairly large angular distance of $\sim$$2.5^{\mathrm{o}}$ from the target. For this reason, on March 19, 2013, we performed VLBI observations of the fly-by event field with the EVN (experiment code ET027). We were able to identify three new sources, one of which -- `CAL5' or \object{J1243$-$0218} (0.133 Jy total flux density; see Fig. \ref{gr035:setup}) -- appeared to be suitable both as a secondary calibrator and as an arc stability test source to ensure the quality of the final astrometrical solution for MEX.

\begin{figure*}
  \centering
  \includegraphics[width=250pt]{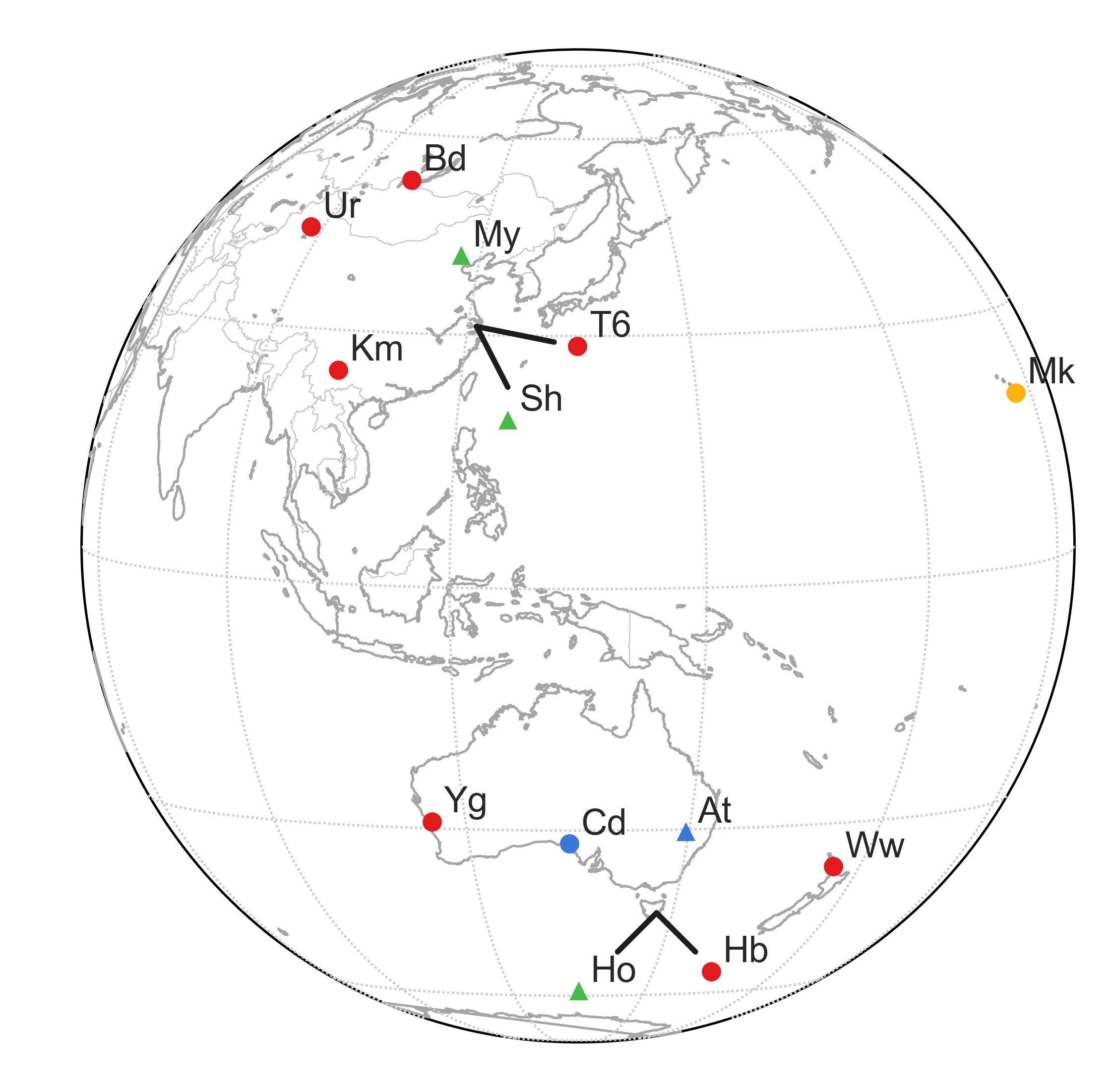}
  \includegraphics[width=250pt]{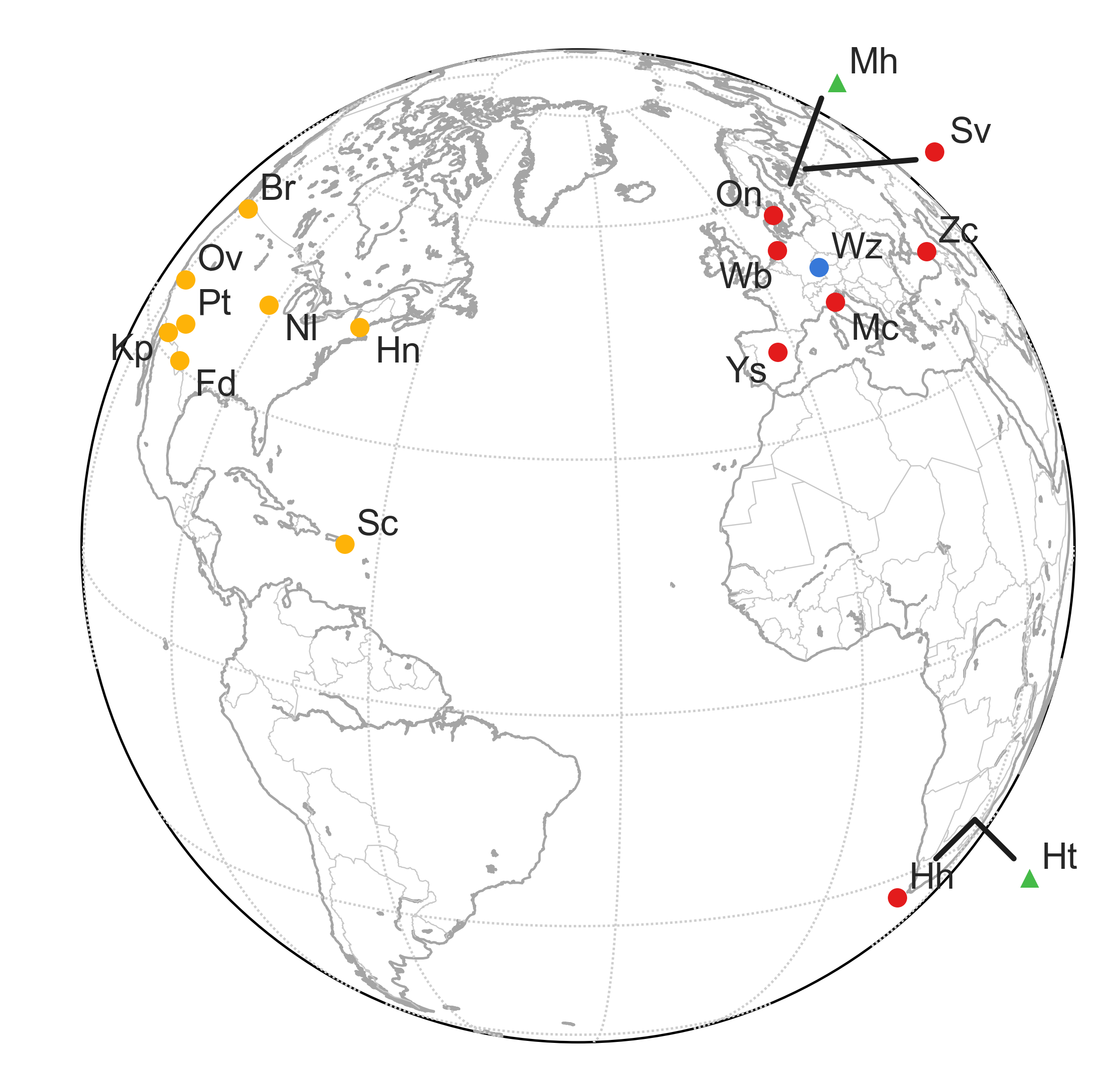}
  \caption{Experiment GR035: participating telescopes are denoted by their two-letter codes. Stations depicted with a triangle were engaged in the Doppler part of the experiment, with a circle in the phase-referencing VLBI part. Station colour denotes frequency set-up used: red -- 8332 -- 8444 MHz, 8 channels of 16 MHz USB (Upper Side Band); blue -- 8396 -- 8444 MHz, 4 channels of 16 MHz USB; orange -- 8380 -- 8428 MHz, 4 channels of 16 MHz USB; green -- 2289 -- 2305 / 8396 -- 8412 MHz, 4 channels of 16 MHz USB.}
  \label{gr035-worldmap}
\end{figure*}

The observations of the fly-by event were carried out within the global VLBI experiment GR035. During the run on December 28-29, 2013, we observed three consecutive revolutions of MEX around Mars, each 7 hours long, in order to provide a better coverage for further orbit reconstruction. The total duration of the experiment was 25 hours. Telescopes that took part in the observations\footnote{Additional information about particular antennas can be found by the two-letter codes shown in Fig. \ref{gr035-worldmap} in the databases of, e.g. the International VLBI Service at \url{ftp://ivscc.gsfc.nasa.gov/pub/control/ns-codes.txt} and the European VLBI Network at \url{http://www.evlbi.org/user_guide/EVNstatus.txt}} were split into two sub-arrays. The antennas in the first sub-array (see Fig. \ref{gr035-worldmap}, stations depicted with triangles) observed in Doppler mode with a dual S/X-band (2/8.4 GHz) frequency set-up with long scans (20-minutes) to minimise the Doppler frequency detection noise.
The second sub-array antennas (see Fig. \ref{gr035-worldmap}, stations depicted with circles) used an X-band (8.4 GHz) set-up and operated in a phase-referencing VLBI mode with short scans \citep{1999A&A...348..381R}, alternating between the target and the phase calibrators.
The frequency set-up used by each of the stations is shown in Fig. \ref{gr035-worldmap}. 
Standard VLBI recording equipment was used at all stations. Strong sources \object{M87} and \object{J1222+0413} with total X-band flux densities of 1.968 and 0.701 Jy, respectively \citep{RFC}, were used as fringe\footnote{The response of a VLBI system.} finders.

\begin{figure}
   \centering
   \includegraphics[width=250pt,clip]{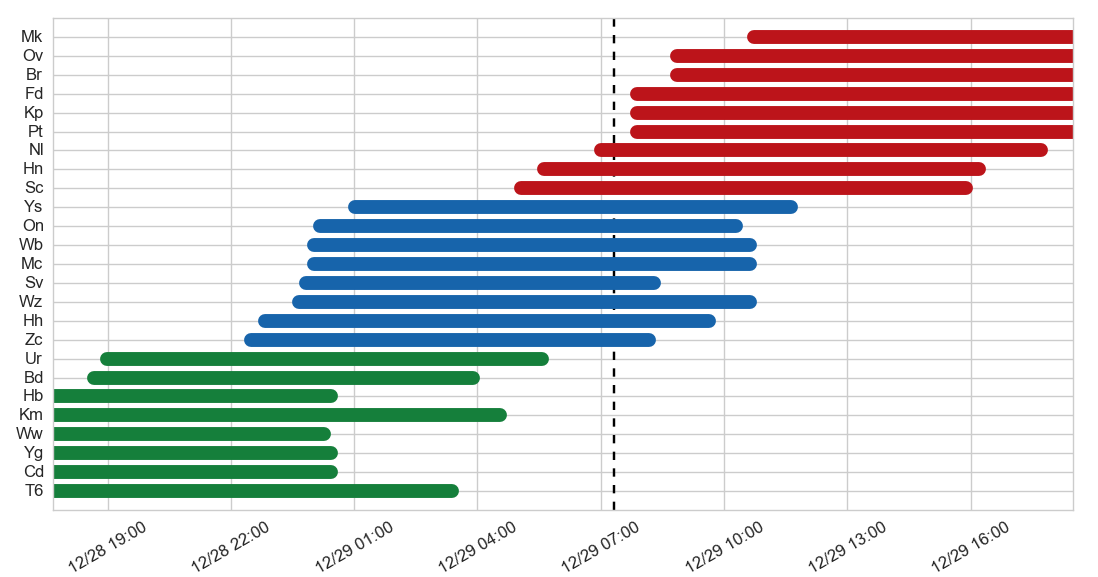}
   \caption{Observational time ranges (UTC) of the telescopes that participated in the phase-referencing VLBI part. Green indicates the Australian, New Zealand and eastern EVN stations, blue the western EVN stations, and red the VLBA stations. Station codes as in Fig. \ref{gr035-worldmap}. The dashed black vertical line denotes the time of the fly-by event as seen from the Earth.}
   \label{GR035:tranges}
    \end{figure}

The Australian, New Zealand and eastern EVN stations began the tracking, then subsequently the western EVN and VLBA stations stepped in. This is illustrated in Fig. \ref{GR035:tranges}.

\section{Doppler and VLBI data processing pipeline and results}

   \begin{figure}
   \centering
   \includegraphics[width=230pt,clip]{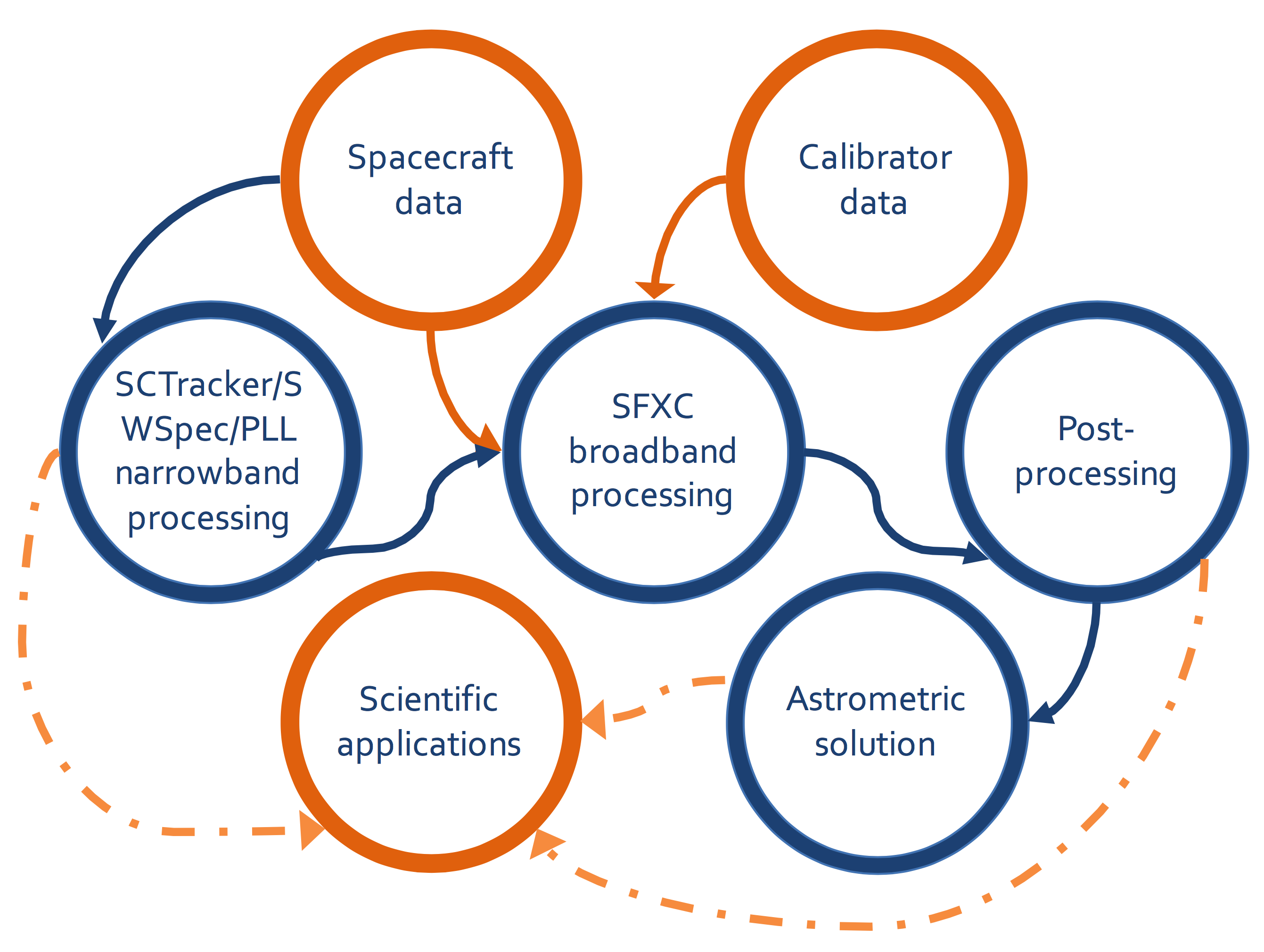}
   \caption{Generic PRIDE data flow and processing pipeline. Spacecraft Doppler and delay observables obtained, respectively, at the narrowband and post-processing steps, alongside the final astrometric solution, that can be used for a variety of scientific applications \citep{2012A&A...541A..43D}.}
   \label{pride:pipeline}
    \end{figure}
    
   \begin{figure}
   \centering
   \includegraphics[width=260pt,clip]{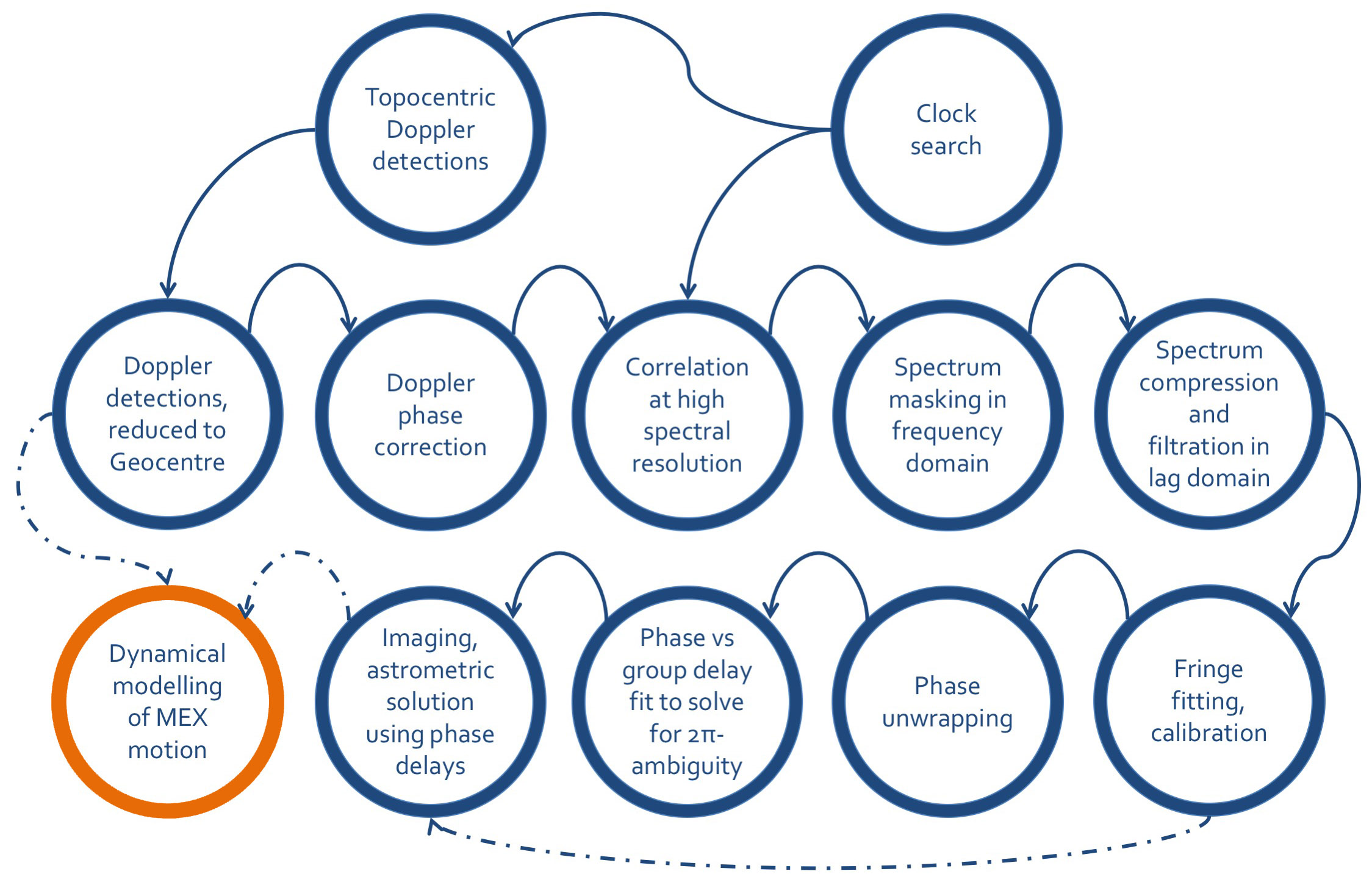}
   \caption{GR035 data processing pipeline. The positional measurements and the Doppler detections are fed into a dynamical model of MEX motion (Rosenblatt et al. 2016, in prep.).}
   \label{gr035:pipeline}
    \end{figure}
    
We have advanced the generic spacecraft data processing pipeline developed within the scope of PRIDE (outlined in Fig. \ref{pride:pipeline}) to achieve a very high precision of MEX positional estimation, which is shown schematically in Fig. \ref{gr035:pipeline}.

First, we performed narrowband processing of the single-dish open-loop data collected by all of the stations. We used the SWSpec/SCtracker/dPLL\footnote{Wagner,~J., Molera Calv\'{e}s,~G., and Pogrebenko,~S.V. 2009-2014, Mets\"{a}hovi Software Spectrometer and Spacecraft Tracking tools, Software Release, GNU GPL, \url{http://www.metsahovi.fi/en/vlbi/spec/index}} \citep{GuifrePhD, 2014A&A...564A...4M} software package to obtain the topocentric Doppler detections. SWSpec extracts the raw data from the channel where the spacecraft carrier signal is expected to be recorded. Then it performs a window-overlapped add (WOLA) discrete Fourier transform (DTF) and time integration over the obtained spectra. The result is an initial estimate of the spacecraft carrier tone along the scan. The moving phase of the spacecraft carrier tone throughout the scan is modelled by performing an $n$-order frequency polynomial fit. SCtracker uses this initial fit to stop the phase of the carrier tone, allowing subsequent tracking, filtering and extraction of the carrier tone in narrower bands (from the initial 16 MHz channel bandwidth down to a 2 kHz bandwidth) using a second-order WOLA DFT-based algorithm of the Hilbert transform approximation. The Digital Phase-Locked-Loop (dPLL) performs high precision reiterations of the previous steps -- time-integration of the overlapped spectra, phase polynomial fitting, and phase-stopping correction -- on the 2~kHz bandwidth signal, using 20000 FFT points and 10-second integration time. The output of the dPLL is the filtered down-converted signal and the final residual phase in the stopped band with respect to the initial phase polynomial fit. The bandwith of the output detections is $20$ Hz with a frequency spectral resolution of $2$~mHz. The Doppler observable is obtained by adding the base frequency of the selected channel to the 10-second averaged carrier tone frequencies retrieved by the dPLL.

   \begin{figure}
   \centering
   \includegraphics[width=260pt,clip]{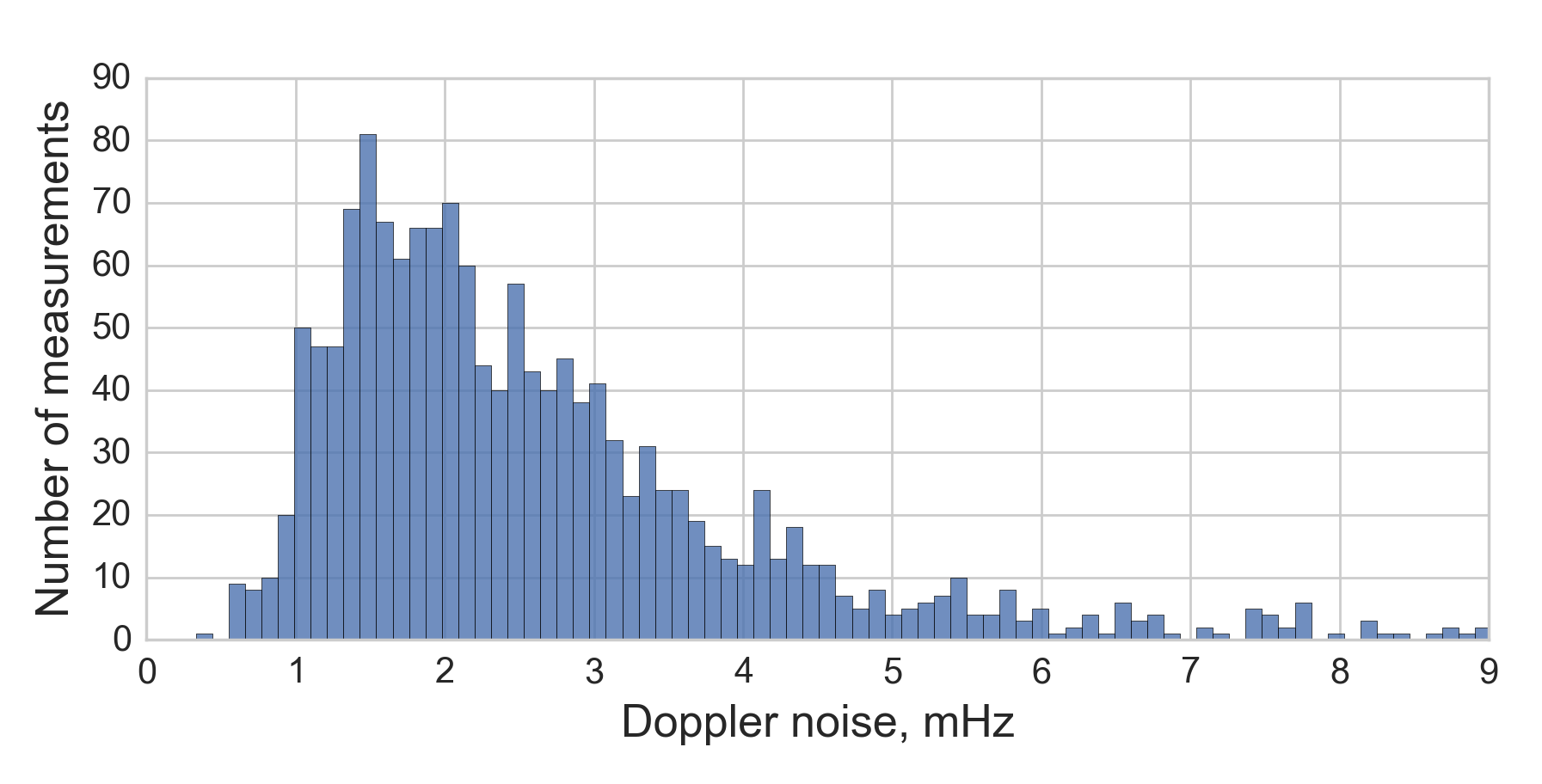}
   \caption{Doppler detection noise in mHz at X-band. Topocentric detections with a 10-second integration time for each station were differenced with predicted values, and the standard deviation of the result was calculated for each 2-minute scan providing what is referred to as the measurements in the histogram.}
   \label{gr035:dopnoise}
    \end{figure}

During this experiment, three transmitting stations provided 24-hour coverage: the 35-metre ESTRACK station New Norcia (NNO) in Australia and the 70-metre Deep Space Stations 63 (DSS-63) in Robledo (Spain) and 14 (DSS-14) in Goldstone (CA, USA). In order to estimate the Doppler noise $f_{n}$, the topocentric detections with a 10-second integration time for each station were differenced with predicted three-way Doppler values, and the standard deviation of the result was calculated for each 2-minute scan (see the histogram in Fig \ref{gr035:dopnoise}). The resulting mean value of $f_{n}$ is 2.5 mHz, median -\,2.2 mHz, and mode (maximum of a fitted log-normal distribution) -\,1.7 mHz. The mode value translates to $0.5 \cdot c \cdot (f_n/f_0) = 30~\mathrm{\mu}$m/s in linear measure for the three-way Doppler, where $f_0=8.4$ GHz, and $c$ is the speed of light in a vacuum. This is comparable to the precision of the Doppler detections provided by the DSN and ESOC \citep[see e.g.][]{1992JGR....97.7759T, 2004ESASP.548..387B}.

In order to process the VLBI data, streams from each station must be synchronised with a common base, usually the International Atomic Time TAI. The behaviour of station clocks is regularly checked against the GPS\footnote{Global Positioning System} time scale tied to the TAI time. We examined these time series and chose Medicina (station code Mc) as the absolute reference station for the current experiment, which appeared to have the best long-term stability and the smallest absolute clock rate value around the date of the experiment. Clock parameters of the rest of the stations were referenced to Medicina using the fringe finder data\footnote{This procedure is usually referred to as the `clock search'}.

   \begin{figure*}
   \centering
   \includegraphics[width=360pt,clip]{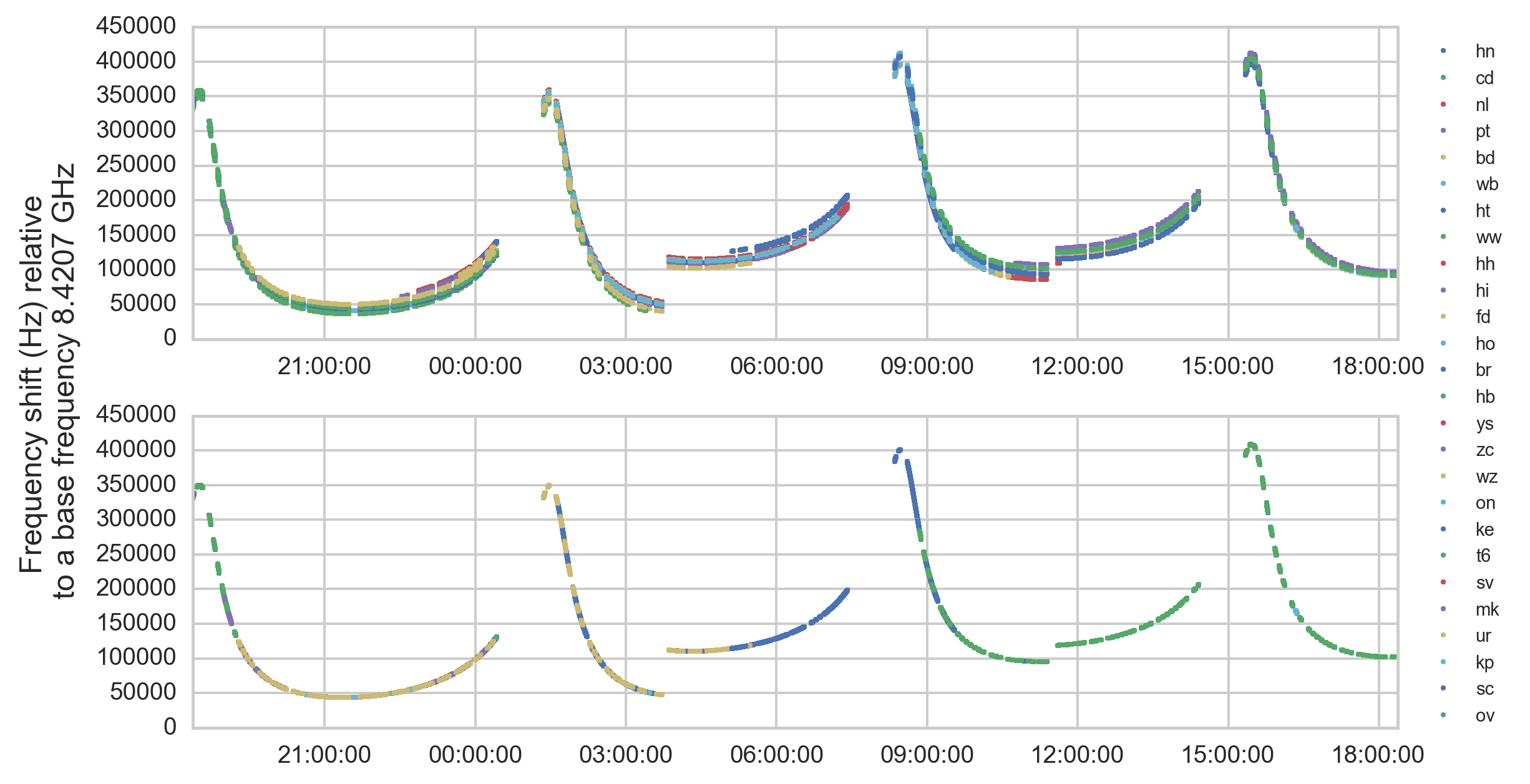}
   \caption{Topocentric frequency detections (top, Hz) and frequency detections reduced to a common phase centre - geocentre (bottom, Hz).
Jumps in frequency at 03:30 and 11:30 UT are due to the uplink frequency changes at transmitting ground stations. Mean geocentric frequency was converted to phase
and applied to the spacecraft signal at correlation to avoid a frequency smearing at high spectral resolution. Station two-letter codes as in Fig. \ref{gr035-worldmap}.}
   \label{gr035:fgc}
    \end{figure*}

The spacecraft cross-correlation spectrum is smeared owing to the intrinsic change of the frequency (emitted by a spacecraft as it retransmits the signal in the two-/three-way Doppler regime) caused by a change in the relative velocity. To mitigate this frequency smearing, a Doppler phase correction must be applied to the spacecraft data. In VLBI, data from individual telescopes are reduced to a common phase centre, usually the geocentre. Therefore, to compute an empirical Doppler phase correction, we first reduced all the topocentric frequency detections $f_\mathrm{tc}(t)$ to the geocentre (see Fig. \ref{gr035:fgc}) using the equation
\begin{equation}
f_\mathrm{gc}(t) = f_\mathrm{tc}(t -\tau_\mathrm{gc} )\cdot(1-\frac{\mathrm{d}\tau_\mathrm{gc}}{\mathrm{d}t}),
\label{eq1}
\end{equation}
where t is UTC time, and $\tau_\mathrm{gc}$ and $\mathrm{d}\tau_\mathrm{gc} / \mathrm{d}t$ are the total near-field VLBI signal delays and delay rates with respect to the geocentre.
The resulting geocentric frequencies (consistent with each other and with the geocentric frequency prediction at a sub-mHz level) are subsequently averaged and integrated into phases on a per scan basis. This way, phases are reset to zero at the beginning of each scan. This is done to avoid numerical errors associated with a fairly wide dynamical range of changes in the phase.

   \begin{figure}
   \centering
   \includegraphics[width=250pt,clip]{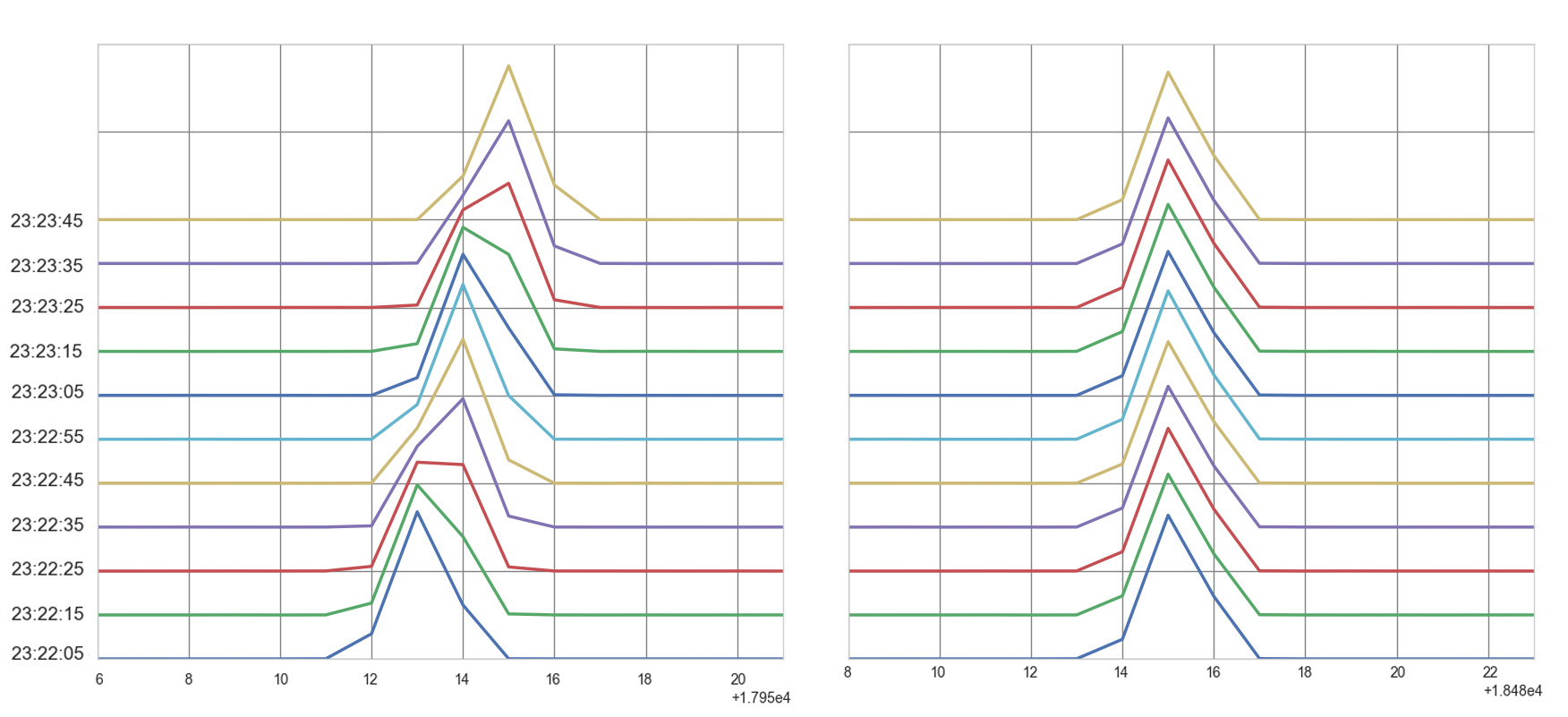}
   \caption{Zoom into the carrier line without (left panel) and with the Doppler phase correction (right panel). 10 sec integration time, $2^{15}=32768$ points spectral resolution, baseline Hh-Ww, scan 135. 23:22 -- 23:24 UTC, December 28, 2013.}
   \label{gr035:phasecor}
    \end{figure}

   \begin{figure}
   \centering
   \includegraphics[width=250pt,clip]{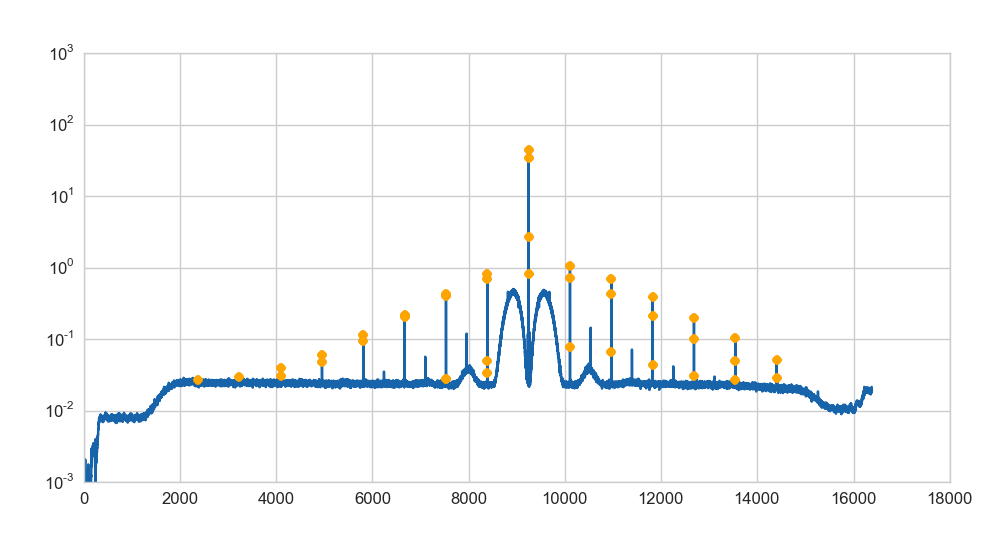}
   \includegraphics[width=250pt,clip]{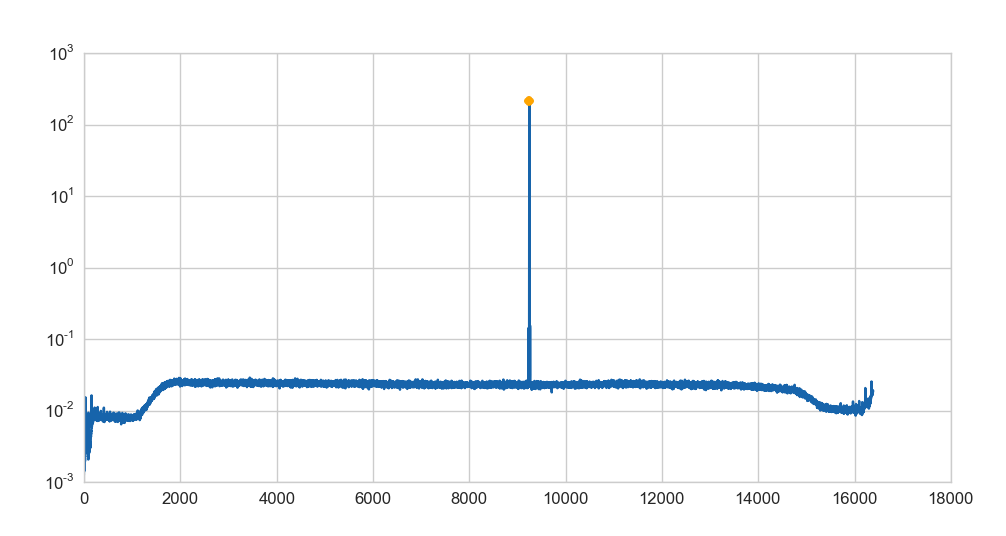}
   \caption{Averaged amplitude spectrum after applying the Doppler phase correction in arbitrary units, $2^{14}=16384$ points spectral resolution, baseline T6-Sv, scan 209 (top) and 191 (bottom). Only the carrier line was present in the spectrum in the second case, as was the case for $\sim 50\%$ of the time during GR035. The spectral mask is shown in orange dots.}
   \label{gr035:specmask}
    \end{figure}

   \begin{figure}
   \centering
   \includegraphics[width=260pt,clip]{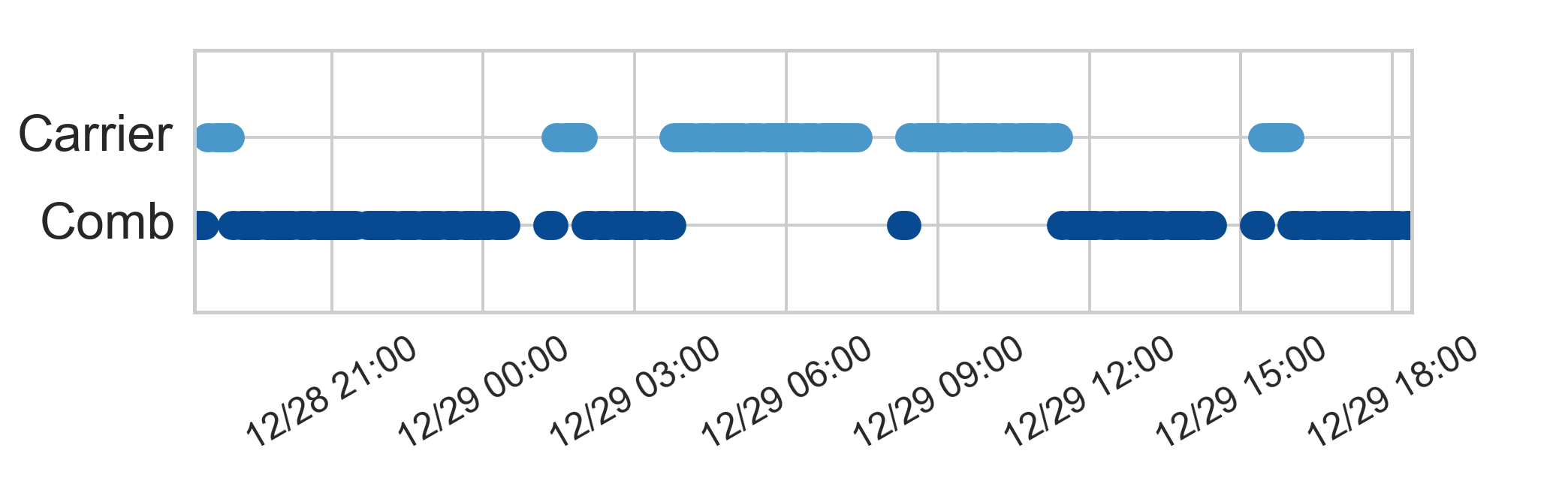}
   \caption{MEX signal spectrum types. `Comb' denotes the time intervals when ranging tones were present in the spectrum; `Carrier' when only the carrier was present in the spectrum.}
   \label{gr035:spectypes}
    \end{figure}

   \begin{figure}
   \centering
   \includegraphics[width=250pt,clip]{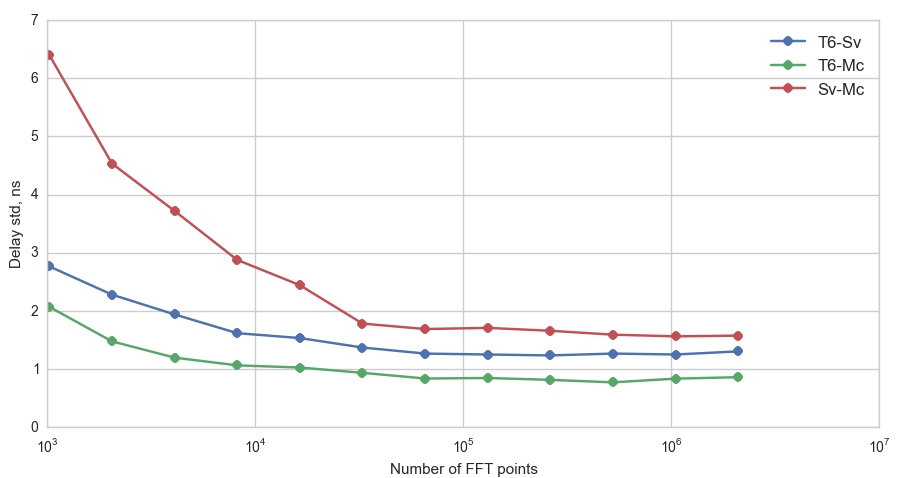}
   \caption{Optimal FFT size to perform signal filtration as characterised by the standard deviation of the group delay estimates obtained after fringe-fitting as a function of  the spectral resolution used at correlation. Spectral resolution ranges from $2^{10}=1024$ to $2^{21}=2097152$ points. Baselines T6-Sv, T6-Mc, Sv-Mc.}
   \label{gr035:fftsize}
    \end{figure}

The resulting phase correction is applied at the next processing step, the broadband correlation with the EVN software correlator at JIVE, SFXC \citep{2015ExA...tmp....9K}. For the correlation of both the calibrator and the spacecraft data, we used the signal delay models described in \citet{2012A&A...541A..43D}. These models have been implemented in a software package $pypride$ (PYthon tools for PRIDE), whose output is compatible with the SFXC. The package is mostly written in the Python programming language with an extensive use of modules providing JIT-compilation to boost performance. The most computationally expensive sub-routines are written in Fortran. Most of the tasks are automated and parallelised.
The package $pypride$ calculates VLBI delays and the $uvw$-projections of baselines/Jacobians for far- and near-field sources and for space VLBI \citep[for details, see][]{2012A&A...541A..43D, DuevRA}. The software can also be used to calculate Doppler frequency shift predictions for spacecraft observations. 

   \begin{figure*}
   \centering
   \includegraphics[width=460pt,clip]{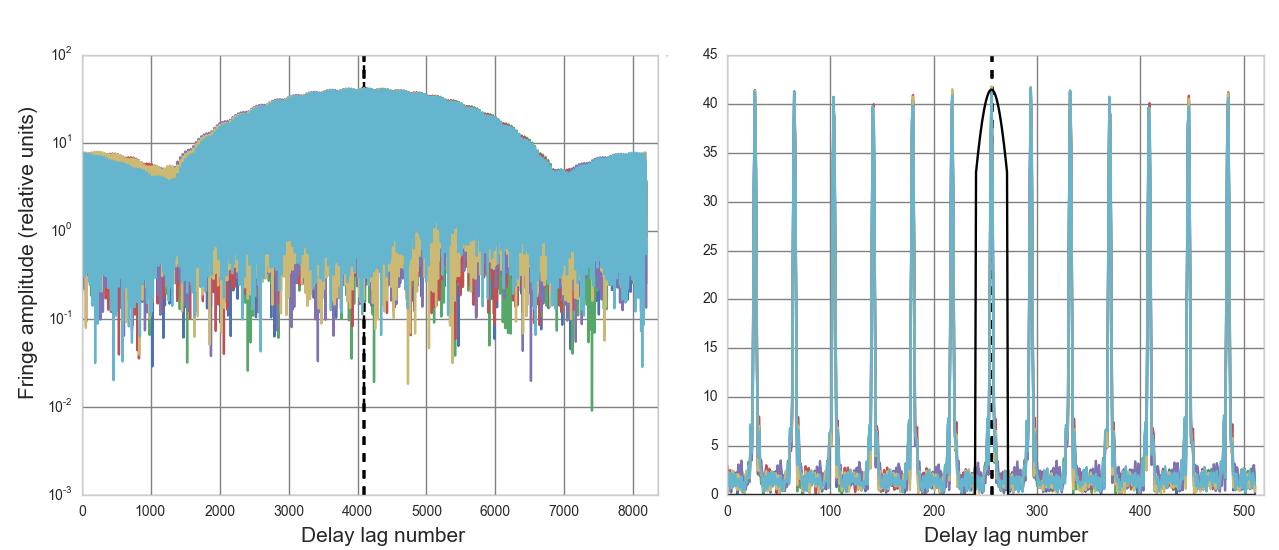}
   \caption{Spectrum compression from 4096 (left panel) down to 256 (right panel) spectral points, and filtration in lag domain. The solid black line in the right panel shows the output filter profile. One-second integrated spectra are shown. Scan 209, baseline T6-Sv. The dashed black lines denote the central lag.}
   \label{gr035:compression}
    \end{figure*}

Changes in spacecraft signal spectrum over time are due to different transmission modes used during a communication session (see Fig. \ref{gr035:spectypes}). A straightforward approach to the correlation of such data at typical resolutions used in VLBI works only if the so-called data-bands are present in the spacecraft spectrum (see Fig. \ref{gr035:specmask}, left -- characteristic `bumps' around the carrier and the first sub-carriers). However, this approach effectively narrows the bandwidth by a factor of $\sim3$, which results in higher noise in the group delay estimates at the next processing step. More importantly, however, these spectral features are not constant over time and may completely disappear for extended intervals. To overcome these difficulties, we realised an approach that makes use of the ranging tones (sub-carriers) present in the S/C spectrum most of the time; the phases of these tones are directly related to the carrier phase as they are all synthesised from the same reference signal. This allows most of the available bandwidth to be used. The Doppler phase correction stops the cross-correlation spectrum drift (see Fig. \ref{gr035:phasecor}). Individual sub-carrier lines are clearly seen only at a sufficiently large spectral resolution owing to their intrinsic narrowness. Therefore, we first correlated data on several baselines with larger telescopes of the array using a very high spectral resolution and derived a spectral mask leaving one to four spectral points to a line depending on its width at that particular resolution. For an initial mask approximation, we used a peak identification approach incorporating continuous wavelet transform-based pattern matching \citep{Du01092006}, after which the resulting mask was inspected and corrected by hand if necessary. To derive the optimal spectral resolution, we tried fringe-fitting the results of correlation at different resolutions on several baselines (see Fig. \ref{gr035:fftsize} for examples). The standard deviation of the 1-second integrated group delay estimates suggested an optimal spectral resolution value of $2^{19}=524288$ points. Presumably, numerical effects come into play at higher resolutions, preventing a further increase in precision. Finally, the amplitudes of individual filtered lines are normalised to unity, while the phases are kept intact. This provides additional improvement in the precision of group delay estimation by making the fringes more pronounced in the lag domain.

We need to point out that for $\sim$$50\%$ of the time during GR035, only the carrier line was present in the spacecraft spectrum (see Fig. \ref{gr035:specmask}, bottom; the mask here consists of a few points around the carrier). In this case, preserving this only feature in the spectrum does not allow group delay estimation owing to an extremely narrow effective bandwidth; however the phase may still be accurately extracted and used.

Correlation at such a high spectral resolution results in a massive amount of data. Therefore, in order to reduce the latter to a manageable level, the resulting spectra are compressed to a resolution of 256 spectral points. This is achieved by transforming the spectra into the lag domain, where 512 points around the central lag are cut out. These spectra are then transformed back into the frequency domain (see Fig. \ref{gr035:compression}). The secondary peaks seen in Fig. \ref{gr035:compression} (right) around the central (true) fringe are the result of the compression operation at the previous stage, which is equivalent to convolving the comb-like spectra\footnote{A spectrum with filtered and normalised spectral lines resembles a comb.} in frequency domain with a Fourier image of a set of rectangular windows, and can easily confuse the fringe fitting algorithm. However, after performing the clock search, we know that the fringe maximum in the lag domain lies within several lags around the zeroth lag, and so we can eliminate these `out-of-the-maximum' peaks by applying a squared cosine-window filter (shown by a black line in Fig. \ref{gr035:compression}, right). Our tests have shown that fringe fitting the output of this compression procedure yields the same precision of group delay estimates as in the case of the original non-compressed data.

The spectrum filtration and compression described above were first implemented in Python and thoroughly tested before being incorporated into the SFXC correlator. This implementation is integrated in the SFXC standard spectral averaging code and allows the arbitrary spectral and window filters to be specified as appropriately sized vectors.

The output cross-correlation spectra in the SFXC correlator-specific format were converted into the Measurement Set format \citep{MS} and into the FITS IDI files for further processing. 

\begin{figure*}
  \centering
  \includegraphics[height=160pt]{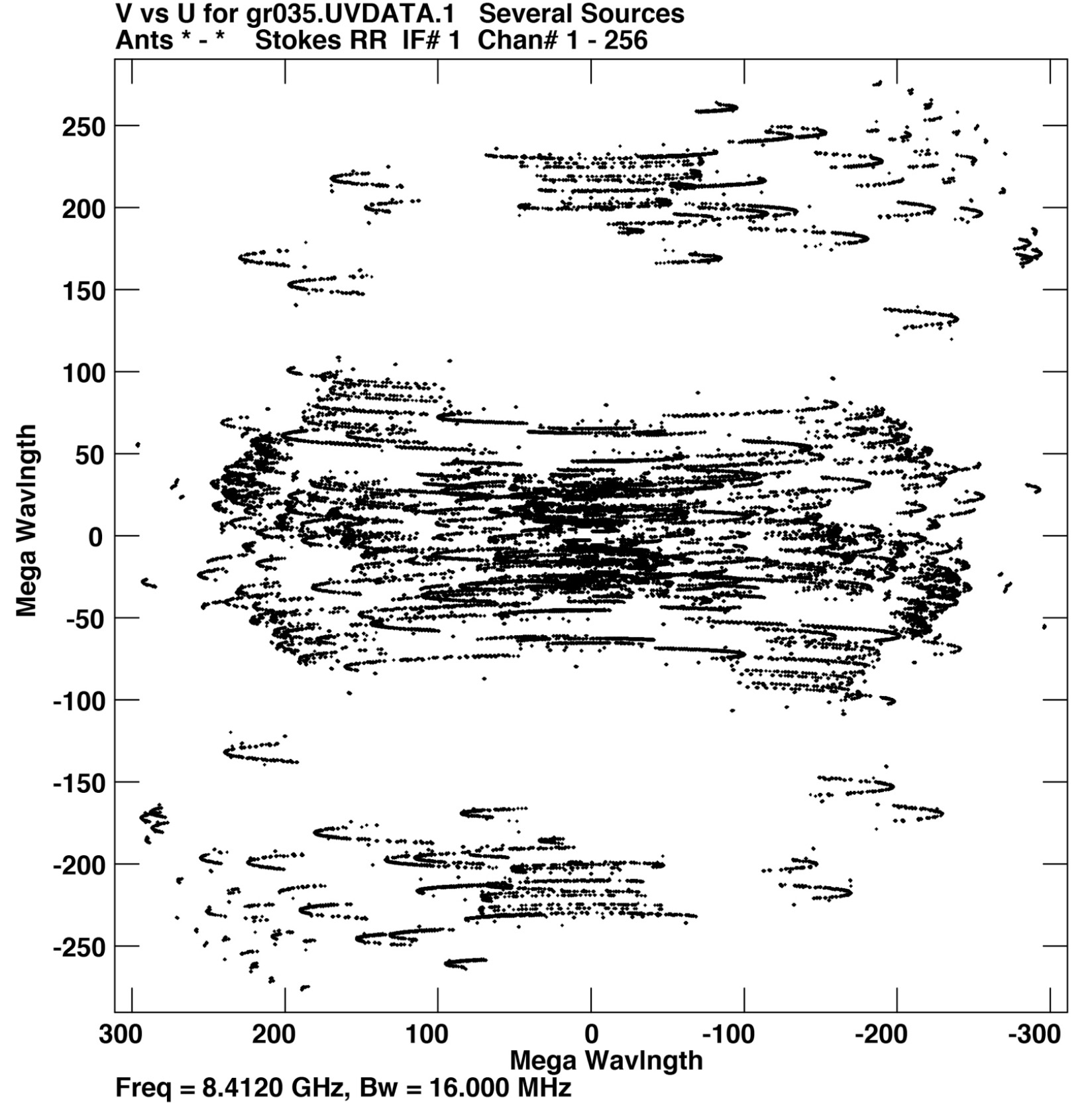}
  \includegraphics[height=160pt]{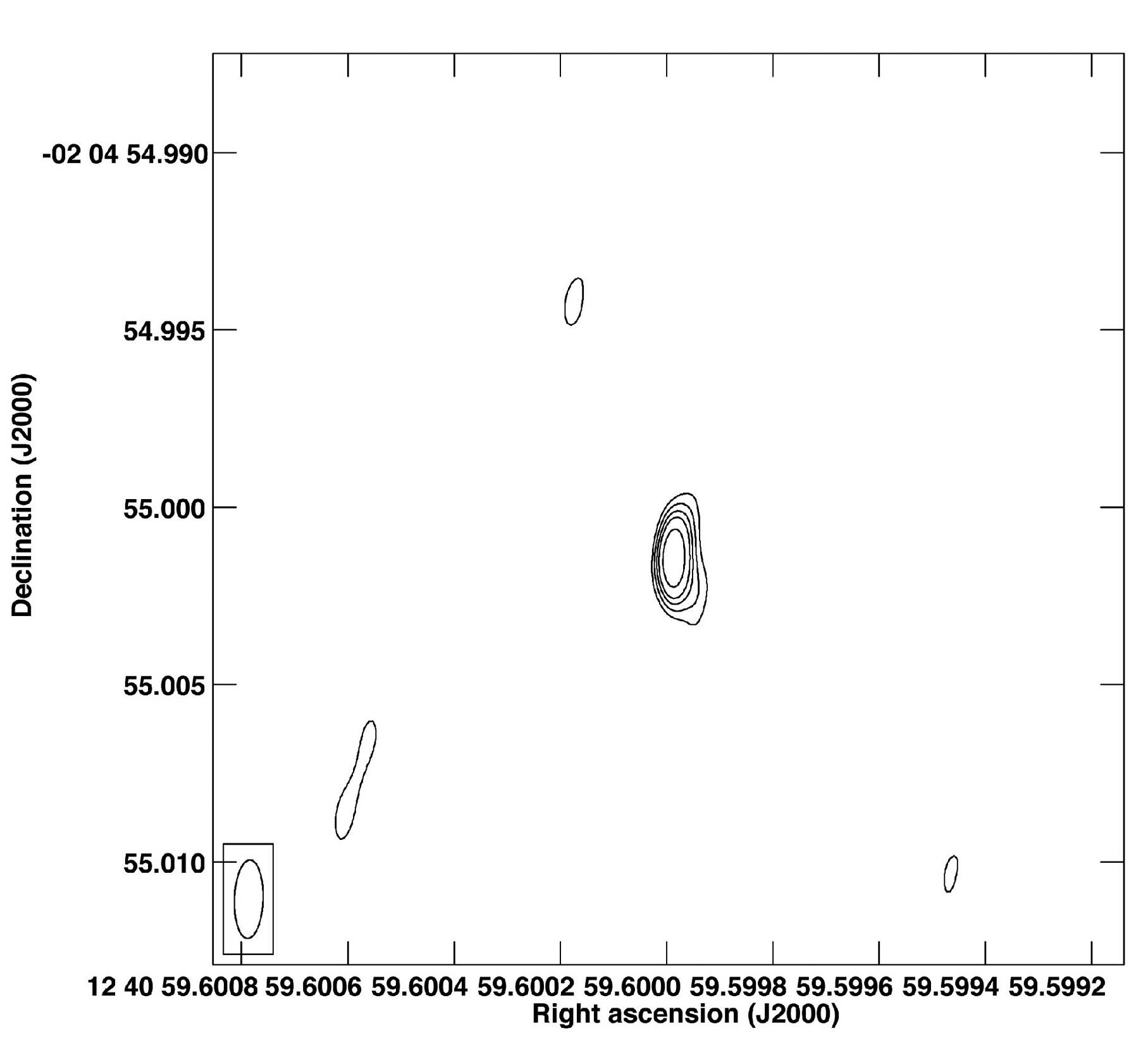}
  \includegraphics[height=160pt]{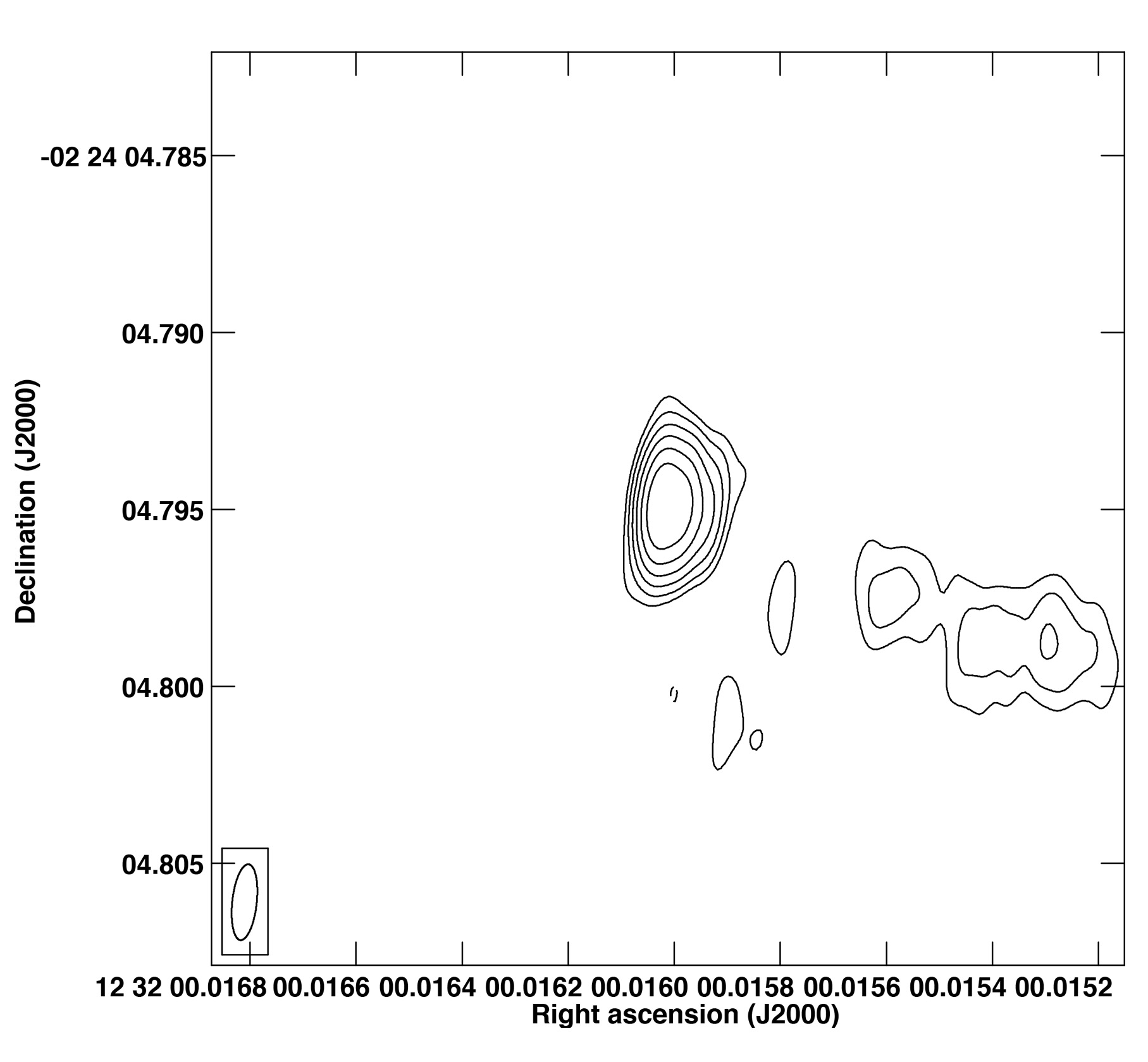}
 \caption{Left panel shows full time range $uv$-coverage for all sources. Middle panel -- CLEAN'ed map of MEX integrated over the time range from 29/12 11:30--13:30 UTC. Right panel -- self-calibrated CLEAN'ed map of the primary calibrator source \object{J1232$-$0224} to the same scale and integrated over the same time range.}
  \label{gr035-uv-img}
\end{figure*}

To derive the displacements of MEX from its a priori position in the post processing analysis of the data, we employed two different independent approaches: imaging and solving the astrometric measurement equation.

The first approach was realised employing a commonly used VLBI data reduction package AIPS \citep{2003ASSL..285..109G}. The FITS-files are loaded into the AIPS file system using a task\footnote{In the AIPS environment, separate sub-programs are called `tasks'.} FITLD. After a preliminary data inspection and editing, initial calibration is applied. This includes bandpass (AIPS task BPASS) calibration using the source \object{J1222+0413} and antenna calibration (task ANTAB). For the stations that did not provide system temperature measurements during the experiment, nominal values were used to calibrate the antenna. To correct the delays and rates of the phase referencing calibrator \object{J1232$-$0224}, fringe fitting is performed with the task FRING. Next, a procedure called `self-calibration' is applied to the calibrator using the CALIB task. During self-calibration, phase corrections for the antennas are calculated based on a model of the source. We started with a point-source model. Then a CLEAN'ed \citep{1974A&AS...15..417H} map of the calibrator is produced using the AIPS task IMAGR. The resulting map is used as a new model. An example CLEAN'ed self-calibrated map of the source \object{J1232$-$0224} is show in Fig. \ref{gr035-uv-img}, right panel. When we are satisfied with the map and with the calibration after a number of iterations of IMAGR and CALIB, the resulting phase corrections are applied to the spacecraft. At this stage it is possible to make an image of the spacecraft. The process of self-calibration fixes the centre of the map to the nominal a priori position\footnote{For 80\% of the calibrators, the position is known to an accuracy better than 3~mas. According to \citet{2016arXiv160507036S}, 1167 calibrators were known within $7.5^\circ$ of the ecliptic band by May 2016, and their number is growing. The median accuracy of their positions is 0.45~mas. A source is considered a calibrator if its median correlated flux density at baselines longer than 5000 km is above 30~mJy at 8 GHz. A dedicated observing program for improving positions of all known calibrators to a level of 0.3~mas is underway \citep{2016arXiv160507036S}. Potentially, the accuracy can be further improved to reach a level of 0.1~mas if the necessary resources are allocated.}. 
However, the position of the spacecraft on the final image will have a shift from the centre of the map (defined by the spacecraft a priori position) due to the errors in the spacecraft ephemeris. The position measured on the map provides the actual coordinates of the spacecraft for each solution interval. The measurements are done with the task JMFIT by fitting a 2D Gaussian to the peak on the CLEAN'ed spacecraft map located within a box set using the image integrated over the full time range for the current sub-array (which efficiently suppresses the side lobes and any quickly changing variations in phase). An example image of MEX is shown in Fig. \ref{gr035-uv-img} (middle panel).

The described pipeline was automated with a ParselTongue script \citep{2006ASPC..351..497K}. When imaging the scans where only the carrier line was present in the MEX spectrum, we used only a few frequency channels around the carrier in a manner analogous to that used in spectral-line VLBI.

   \begin{figure*}
   \centering
   \includegraphics[width=460pt,clip]{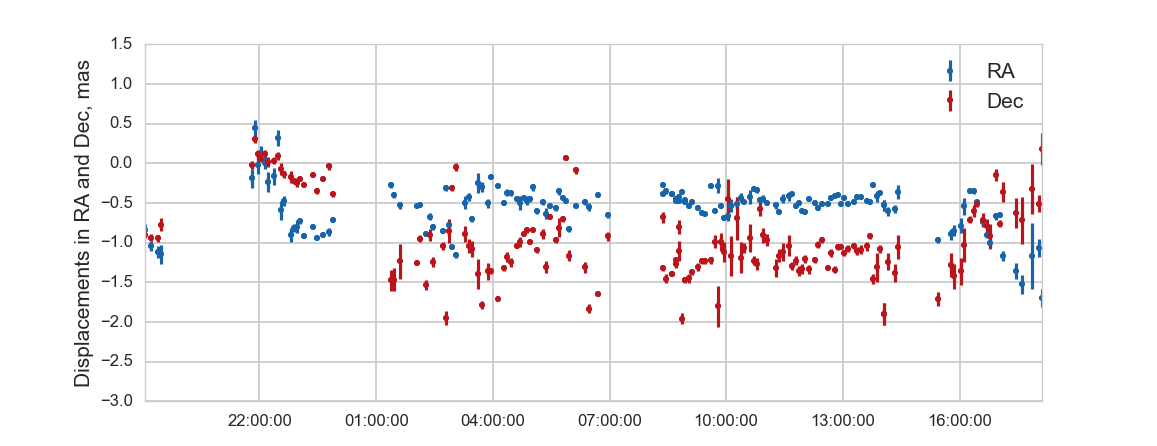}
   \caption{Displacements from the a priori lateral position of MEX as a function of time, measured using the imaging approach. Displacements in Right Ascension (mas) are shown in blue, in Declination (mas) in red. 2-minute integration time; net time on target $\sim5.5^{h}$. December 28-29, 2013. The apparent jump in position near 01:00:00 UT is discussed in section 3.}
   \label{gr035:radec}
    \end{figure*}


The alternative approach we used for estimating the MEX positional displacement (which we realised within the $pypride$ software package) is similar to the approach used in geodetic VLBI and is based on solving the measurement equation for each epoch $t$,

\begin{equation}
\label{ME}
\overrightarrow{\Delta\phi} \ \big|_t = \big( J \cdotp \overrightarrow{\Delta \alpha} \big) \ \big|_t,
\end{equation}
where  $\overrightarrow{\Delta\phi}$ is a vector of differential MEX carrier line phases on baselines, $J$ is a matrix containing near-field analogues of $uv$-projections of baselines \citep[see][]{2012A&A...541A..43D}, and $\overrightarrow{\Delta \alpha}$ is the vector of corrections to the a priori lateral position of the spacecraft. The phases $\overrightarrow{\Delta\phi}$ are subject to a $2\pi$-ambiguity, which means that the corresponding phase delays $\tau_\mathrm{ph} = \varphi / \omega_0$ ($\omega_0 = 2\pi f_0$) may have a bias of several cycles of $\sim$120 ps for observations at X-band. In order to solve for this ambiguity, we employed the following approach.

First, the calibrator data are fringe-fitted and self-calibrated providing calibration group delays ($\tau_\mathrm{gr} = \mathrm{d} \phi / \mathrm{d} \omega, \omega = 2\pi f$) and phases, which are applied to the MEX data. The rms error of the calibrator group delay estimates are used to set baseline weights. Then, the phase of the MEX carrier line is extracted and unwrapped. A naive approach to unwrapping does not work in most cases owing to large gaps and uneven spacing in time, and because the phase slope is not constant. Therefore, we made use of a wrapped Kalman smoother, which is based on a wrapped Kalman filter (WKF) algorithm described in \cite{2013ISPL...20.1257T} (see Appendix \ref{kalman}). The calibrated MEX data are subsequently fringe-fitted. This yields residual group delays for the scans when the sub-carriers are present in its spectrum. 
An SVM\footnote{Support vector machines.}-based unsupervised machine learning algorithm is used to identify and flag outliers in group delay estimates. An optimal fit of phase delays $\tau_\mathrm{ph}$ to group delays is found by minimising the squared error defined as $\sum_{n}\sum_{i}((\tau_\mathrm{ph}[i]+2\pi n)-\tau_\mathrm{gr}[i])^2$ over time periods when both phase and group delays are available with respect to $n \in \mathbb{N}$. This yields the number of phase cycles $n_m$ in question providing a solution to the $2\pi$-ambiguity problem, also for the time intervals when no group delay data are available. If there is such a long gap in time in the data that the Kalman smoothing procedure fails to correctly unwrap the phase, the group delays are automatically split into an appropriate number of clusters using a DBSCAN algorithm \citep{dbscan}, and then the phase delays are fitted to the corresponding clusters. This procedure yields an `unambiguous' phase to be used in the astrometric equation (\ref{ME}) solution. 

A proper transition from group delays\footnote{The spacecraft signal is bandwidth limited to about 12 MHz in our case (see Fig. \ref{gr035:specmask}), which sets a certain limit on the group delay estimation accuracy}, which are commonly used in VLBI astrometry, allows the eq. (\ref{ME}) solution error to be reduced by an order of magnitude.

Equation (\ref{ME}) is in most cases overdetermined, therefore we used the singular value decomposition of the matrix $J$ when solving it. The resulting angular corrections $\overrightarrow{\Delta \alpha}$ are not displacements in Right Ascension and Declination per se (because they are defined for sources at infinity), but the angular displacement of the vector from geocentre to the target at a given epoch \citep[see][]{2012A&A...541A..43D}.


In both approaches described above, the solution interval was set to the MEX scan length (2 minutes in most cases); for each target scan, two adjacent calibrator scans were used to perform calibration. In the processing, we used an elevation cut-off angle of 20$^{\mathrm{o}}$. Three stations were set as reference stations: Bd (from the start of the experiment until $\sim$ December 29, 2013 00:30 UTC), Ys (from $\sim$ 01:00 until $\sim$ 07:30 UTC), and Pt (from $\sim$ 08:00 UTC to the end of the experiment). In the first part of the observations, the largest telescope of that sub-array (see Fig. \ref{GR035:tranges}, stations denoted in green), T6, had to be dropped because of phase stability problems at the station. In addition, most of the stations of the sub-array did not provide satisfactory system temperature T$_{sys}$ measurements, which resulted in poor data calibration and a consequent astrometric solution bifurcation. In Fig. \ref{gr035:radec} (before $\sim$ December 29, 2013 00:30 UTC) the solution nearest to the phase centre is shown. However, the solutions for the second and third sub-arrays with Ys and Pt as reference stations (Fig. \ref{gr035:radec} after $\sim$ December 29, 2013 01:00 UTC) show no bifurcation. Finally, the results of the imaging and `geodetic' approaches are consistent at a level of $\sim10$ micro-arcseconds for the best-calibrated time range (from $\sim$ {December 29, 2013} 01:00 UTC onwards). The median $3\sigma$ formal error values for the full time range are $0.034$ mas for RA and $0.058$ mas for Dec, which translates into $\sim$ 35 and 60 m at the orbit, respectively (MEX was at a distance of $\sim1.4$ AU during the experiment).

We checked the stability of our array of telescopes by performing imaging of the source CAL5 (\object{J1243$-$0218}) using all available observations and the data of different sub-arrays. The derived astrometric position in all cases appeared to be the same as in the experiment ET027 with an uncertainty of less than 0.1 mas. At the same level of uncertainty, the coordinates are consistent with the Radio Fundamental Catalogue \citep{RFC} values (RA $12^{h}43^{m}52.4878640^{s}$; Dec $-02^{d}18'38.401056''$).

\section{Conclusions and outlook}
In this work, we were able to measure the lateral position and radial Doppler of the MEX spacecraft with a precision of  about 50$\,$m and $30\,\mathrm{\mu}$m/s, respectively. This is comparable to what has been reported for other spacecraft \citep[e.g.][]{2015AJ....149...28J}. These measurements are used by our collaborators from the Royal observatory of Belgium (ROB) and the French National Centre for Space Studies (CNES) in the dynamical modelling of Mars Express motion aimed at estimating the geophysical parameters of Phobos (see Rosenblatt et al. in prep.).

The offsets in the estimated MEX RA and Dec of $\sim$1 mas from the a priori orbital position seen in Fig. \ref{gr035:radec}, although comparable to the formal orbit determination (OD) error budget, require further investigation and calibration. Similar results have been reported in the past for other spacecraft \citep[e.g.][]{2005IPNPR.162A...1L}. To calibrate these offsets, observations of multiple spacecraft/calibrator pairs with telescope networks of different configurations will be required. In addition, models of signal delay caused by propagation effects must be improved. These are necessary since the systematic offsets most likely result from a combination of a number of error sources, but are dominated by the uncertainty in the a priori spacecraft position and propagation effects. 

PRIDE was selected by ESA as one of the experiments of its L-class JUpiter ICy moons Explorer mission (JUICE) \citep{2013P&SS...78....1G}. The spacecraft data acquisition, processing, and analysis pipelines developed in this work will create a basis for implementation of PRIDE-JUICE.

\begin{acknowledgements}
We would like to express our gratitude to the anonymous referee. 
The authors acknowledge the EC 7th Framework Programme (FP7/2008-2017) project ESPaCE (grant agreement $\#$263466). Mars Express is a mission of the European Space Agency. Information about Mars Express telecommunication were provided by the Mars Express Project.
The European VLBI Network is a joint facility of independent European, African, Asian, and North American radio astronomy institutes. Scientific results from data presented in this publication are derived from the following EVN project code: GR035. The National Radio Astronomy Observatory is a facility of the National Science Foundation operated under cooperative agreement by Associated Universities, Inc. The Australia Telescope Compact Array is part of the Australia Telescope National Facility which is funded by the Commonwealth of Australia for operation as a National Facility managed by CSIRO. This study made use of data collected through the AuScope initiative. AuScope Ltd is funded under the National Collaborative Research Infrastructure Strategy (NCRIS), an Australian Commonwealth Government Programme.
The authors would like to thank the personnel of the participating stations. R.M.~Campbell, A.~Keimpema, P.~Boven (JIVE), M.~P{\"a}tzold (University of Cologne), B.~H{\"a}usler (University of BW Munich), and D.~Titov (ESA/ESTEC) provided important support to various components of the project. G.~Cim\'{o} acknowledges the Horizon 2020 project ASTERICS. T.~Bocanegra Bahamon acknowledges the NWO--ShAO agreement on collaboration in VLBI. J.~Kania acknowledges the ASTRON/JIVE Summer Studentship programme. P. Rosenblatt is financially supported by the Belgium PRODEX program managed by the European Space Agency in collaboration with the Belgian Federal Science Policy Office.
\end{acknowledgements}

\bibliographystyle{aa} 
\bibliography{ref} 

\vspace{20pt}
\begin{appendix}
\section{Wrapped Kalman smoother}\label{kalman}

The wrapper Kalman filter (WKF) is a Kalman filter for which the filtered state distribution $P_{WN}$ is represented by a wrapped Gaussian \citep{2013ISPL...20.1257T}:

\begin{equation}
P_{WN}(\varphi \ | \ \mu, \sigma^2) = \frac{1}{\sigma\sqrt{2\pi}} \sum_{l=-\infty}^{\infty} \mathrm{exp} \Big[ -\frac{(\varphi-(\mu+2\pi l))^2}{2\sigma^2} \Big], \\ \varphi \in \mathbb{S}^1
\end{equation}

The wrapped Gaussian distribution results from mapping a normally distributed random variable $\gamma \sim \mathcal{N}( \mu, \sigma^2), \varphi \in \mathbb{R}^1$ onto a unit circle $\mathbb{S}^1$:

\begin{equation}
\varphi = \psi(\gamma) = \mathrm{mod}(\gamma+\pi, 2\pi) - \pi
\end{equation}

The filtering algorithm is summarised below:

\begin{align}
\mathrm{\bf Predict:} \nonumber\\
\hat{\bf{z}}_{t}^- = \mathrm{A} \hat{\bf{z}}_{t-1} \nonumber\\
\hat{\bf{z}}_{t}^-[1] = \psi(\hat{\bf{z}}_{t-1}^-[1]) \nonumber\\
\mathrm{\hat{\Sigma}}_{t}^{-} = \mathrm{A} \mathrm{\hat{\Sigma}}_{t-1}^{-} \mathrm{A}^T + \Sigma_v \nonumber\\
\mathrm{\bf Correct:} \nonumber\\
\mathrm{K}_t = \frac{\mathrm{\hat{\Sigma}}_{t}^{-} \mathrm{B}^T}{\mathrm{B} \mathrm{\hat{\Sigma}}_{t}^{-} \mathrm{B}^T + \sigma_w^2} \nonumber\\
~~~~~~~~~~~~~~~~~~~g_t = \sum_{l=-1}^{1} ((\varphi_{t} + 2\pi l) - \hat{\bf{z}}_{t}^-[1])\eta_{t,l} \nonumber\\
\hat{\bf{z}}_{t} = \hat{\bf{z}}_{t}^- + \mathrm{K}_t g_t \nonumber\\
\mathrm{\hat{\Sigma}}_{t} = (\mathrm{I} - \mathrm{K}_t \mathrm{B})\mathrm{\hat{\Sigma}}_{t}^{-}
\end{align}
where $\bf{z}_t = \left[ \begin{array}{c} \varphi_t \\ \dot{\varphi}_t \end{array} \right]$ is the system's state, $\bf{z}_t[1] = \varphi_t$, $\mathrm{A} = \left[ \begin{array}{cc} 1 \ \mathrm{d}t \\ 0 \ 1 \end{array} \right]$ is the linearised state transition matrix, $\mathrm{d}t$ is the time interval between the $(t-1)^{th}$ and $(t)^{th}$ epochs, $\Sigma_v$ is the state covariance matrix, $\Sigma_t$ is the state covariance matrix estimate, $\mathrm{B} = [1 \ 0]$ is the observation matrix, $\sigma_w$ is the observation variance, $\mathrm{I}$ is an identity matrix, and $\eta_{t,l} = \mathcal{N}(\varphi_t + 2\pi l  \ | \   \hat{\bf{z}}_{t}^-[1], \sigma_w^2) / \sum_{m=-\infty}^{\infty} \mathcal{N}(\varphi_t + 2\pi m  \ | \   \hat{\bf{z}}_{t}^-[1], \sigma_w^2) $ represents the probability of a replicate.


First, the filter is run on the phase data turned `backwards' in time running from $t_N$ to $t_1$, where $N$ is the number of data points. The system state $\bf{z}$ is initialised as

\begin{equation}
\bf{z}_0 = \left[ \begin{array}{c} \varphi_N \\ 0 \end{array} \right]
\end{equation}

The output of this filtering procedure at $t_0$ is used as the initial condition for running the Kalman filter `forward' in time. Usually, it is enough to run the filter backward and forward once to get a robust and reliable result, but if several iterations are needed, the output of the forward-run filter at $t_N$ is used to update the initial condition for the backward-run.

\end{appendix}

\end{document}